\documentclass[11pt]{article}
\usepackage[colorlinks=true,citecolor=blue,linkcolor=blue]{hyperref}
\usepackage{multirow}
\usepackage[left=2cm,right=2cm,top=2cm,bottom=2cm]{geometry}
\usepackage{cite}
\usepackage{psfrag}
\usepackage{graphicx}
\usepackage{graphics}
\usepackage{epsfig}
\usepackage{amsmath,amsfonts,amssymb}
\usepackage{epstopdf}
\usepackage{pstcol,pst-fill,pst-grad}
\usepackage{euscript}
\usepackage{pstricks}
\usepackage{wrapfig}
\usepackage{subfig}
\usepackage{slashed}
\usepackage[toc,page]{appendix}
\usepackage{here}

\date{}
\begin{document}
	\title{\vspace{-3cm}
		\hfill\parbox{4cm}{\normalsize \emph{}}\\
		\vspace{1cm}
		{ Laser effect on the final products of $Z$-boson decay}}
	\vspace{2cm}
	
	\author{ M Jakha$^{1}$, S Mouslih$^{2,1}$, S Taj$^1$ and B Manaut$^{1,}$\thanks{Corresponding author, E-mail: b.manaut@usms.ma} \\
		{\it {\small$^1$ Sultan Moulay Slimane University, Polydisciplinary Faculty,}}\\
		{\it {\small Research Team in Theoretical Physics and Materials (RTTPM), Beni Mellal, 23000, Morocco.}}\\			
	{\it {\small$^2$Faculty of Sciences and Techniques, 
		Laboratory of Materials Physics (LMP),
		Beni Mellal, 23000, Morocco.}}		
	}
	\maketitle \setcounter{page}{1}



\date{\today}

\begin{abstract}
Experimentalists have long sought a method that allows them to control as they like the branching ratios of an unstable particle decay and direct some decay to follow one specific desired channel without another. The powerful laser could make this dream come true. In this Letter and within the framework of the standard electroweak model, we investigate theoretically the laser effect on the branching ratios of different $Z$-boson decay modes by calculating analytically the $Z$-boson decay into a pair of fermion-antifermion $(Z\rightarrow f\bar{f})$ in the presence of a circularly polarized electromagnetic field. It is found that, at high intensities, the $Z$-boson could only decay invisibly into neutrinos, and its decay into any other pair of charged fermions becomes impossible due to the increase in the effective mass that fermions acquire inside the electromagnetic field. The influence of the laser frequency and intensity on the lifetime is also included in order to confirm the surprising result obtained for the pion lifetime in a previous paper [ Mouslih S \textit{et al} 2020 \textit{Phys. Rev. D} \textbf{102} 073006 ].
\end{abstract}

\maketitle
\section{Introduction}
How can the electromagnetic (EM) field change the behavior of particles and their properties in the various scattering and decay processes, is a question that was and is still receiving great interest by scientists and researchers \cite{mouslih,liu,keitel}, owing to the tremendous progress made by laser technology in recent times \cite{lasertechn}. Branching ratio (BR) is one of these properties that have not been sufficiently investigated in the decay processes that occur in the presence of the EM field, especially knowing that it is an experimentally measurable quantity. Generally, the laser-assisted decay of unstable particles by virtue of weak interaction is a process that is enjoying a lot of attention and is still nowadays the subject of many scientific papers \cite{decays}. The weak decay processes in the presence of strong laser fields can be classified into two types: first, laser-assisted processes which also exist in the absence of the field but may be altered due to its presence. Second, field-induced processes which can happen just when a background field is available, providing an extra energy reservoir \cite{piazza}. An overview of weak interaction processes in the presence of intense EM fields was given by Kurilin in \cite{kurilin1999}. As an example of laser-induced decay, it was found that the exotic decay $(l^{-}\rightarrow W^{-}\nu_{l})$ which is clearly impossible in vacuum, since the lepton mass is small in comparison with the ones of final particles, becomes possible in the presence of an external EM field \cite{kurilin1999}. In our turn, we study in this Letter the decay of the $Z$-boson in the presence of an EM field, to explore the influence of the latter specifically on the BR of the different decay modes. The $Z$-boson, which is an electrically neutral vector boson and a heavy analog of the photon, holds a special place in physics as the carrier of the weak interaction. It was experimentally discovered almost simultaneously with its charged counterparts $W^{\pm}$ bosons in $1983$ \cite{discovery}. Their experimental discovery was a great victory  of experimental physics and pivotal in the theoretical establishment of the standard model. For this reason, this discovery was crowned with Nobel Prizes for Carlo Rubbia and Simon van der Meer \cite{nobel}. The properties of these particles (mass, decay rates, cross sections of their production) were studied in experiments at the Large Electron-Positron collider (LEP) and Stanford Linear Collider (SLC) and found to be in perfect agreement with the predictions of electroweak theory \cite{elecweaktheory}. In the standard model of electroweak interaction, the $Z$-boson, with mass $M_{Z}=(91.1876\pm0.0021)$ GeV and total decay rate $\Gamma_{Z}=(2.4952\pm0.0023)$ GeV \cite{pdg2020}, can decay into a pair of fermions $( Z\rightarrow f\bar{f})$ excluding a pair of top quarks with a lifetime of about $3\times10^{-25}\text{sec}$. One of the important quantities measured by LEP and SLC colliders is associated with the invisible $Z$-boson decay rate $\Gamma_{\text{inv}}$, which is the partial decay rate to any final states that are rather difficult to detect by standard collider detectors. $\Gamma_{\text{inv}}$ is interpreted, in the standard model, as the partial decay rate with respect to decay into neutrinos $\Gamma_{\text{inv}}(Z\rightarrow \nu\bar{\nu})$. This quantity gives the opportunity to infer the number of light neutrino species \cite{decamp}, and its precise measurement is of particular interest for probing new physics beyond standard model \cite{newphysics,carena}. The measurement of $\Gamma_{\text{inv}}$ at $e^{+}e^{-}$ colliders can be done by using two different methods (direct and indirect). Both methods and their results are explained in some detail in \cite{carena,thesis}. The $Z$-boson decay to a pair of charged fermions in a strong crossed EM field was calculated in $2009$ by Kurilin in \cite{kurilin2009}. He investigated changes in partial decay rate of the $Z$-boson in the presence of strong EM fields and he found that at enormous values of the external-field strength, the t-quark-production process $(Z\rightarrow t\bar{t})$, which is forbidden energetically in the absence of an external field, begins contributing significantly to the total decay rate of the $Z$-boson. Tinsley examined the effect of a magnetic field on the decays of the $Z$ into two massless gauge bosons $(Z\rightarrow gg~~\text{and}~~Z\rightarrow \gamma\gamma)$ which are otherwise forbidden according to the Landau–Yang theorem \cite{landeauyang}. He found that this reactions can be allowed by the existence of a magnetic field \cite{tinsley2001,angom}. The goal of this Letter is to present what effects an EM field has on the BR of the different $Z$-boson decay modes.  We assume the laser to be turned on adiabatically for a period of time adequately longer than the laser-free $Z$-boson lifetime, and the laser intensity utilized is taken so that it does not permit pair creation \cite{paircreation}. We affirm here, as theoretical physicists, that the basis upon which the results obtained are based is purely theoretical; believing that these results may pave the way and offer wider scope for possible experiments in the future. In this work, natural units $c=\hbar=1$ are used throughout.
\section{Theoretical calculation}
We consider the decay of a $Z$-boson into a pair of fermions,
\begin{equation}\label{process}
 Z(q)\longrightarrow f(p_{1})+\bar{f}(p_{2}),~~~(f=l,u,c,d,s,b)
\end{equation}
where $f$ can be a charged lepton or neutrino, as well as one of the two up-quarks ($u$, $c$) or one of the down-quarks ($d$, $s$, $b$), and the arguments are our labels for the associated momenta. We assume that this decay occurs in the presence of a circularly polarized monochromatic laser field, which is described by the following classical four-potential
\begin{align}\label{potential}
A^{\mu}(\phi)=a^{\mu}_{1}\cos(\phi)+a^{\mu}_{2}\sin(\phi),~~~\phi=(k.x),
\end{align}
where $k=(\omega,\textbf{k})$ is the wave 4-vector $(k^{2}=0)$, $\phi$ is the phase of the laser field and $\omega$ its frequency. The polarization 4-vectors $a^{\mu}_{1}=|\text{\textbf{a}}|(0,1,0,0)$ and $a^{\mu}_{2}=|\text{\textbf{a}}|(0,0,1,0)$ are equal in magnitude and orthogonal, which implies $(a_{1}.a_{2})=0$ and $a_{1}^{2}=a_{2}^{2}=a^{2}=-|\text{\textbf{a}}|^{2}=-(\mathcal{E}_{0}/\omega)^{2}$ where $\mathcal{E}_{0}$ is the amplitude of the laser's electric field. We shall assume that the Lorentz gauge condition is applied to the four-potential, so that $k_{\mu}A^{\mu}=0$, which implies $(k.a_{1})=(k.a_{2})=0$, meaning that the wave vector $\textbf{k}$ is chosen to be along the $z$-axis. \\
The lowest-order scattering S-matrix element for the laser-assisted $Z$ decay reads \cite{greiner}
\begin{equation}\label{smatrix}
\begin{split}
S_{fi}(Z\rightarrow f\bar{f})=\dfrac{-ig}{4\cos(\theta_{W})}\int d^{4}x\overline{\psi}_{f}(x)\gamma^{\mu}(g_{V}-g_{A}\gamma_{5})\psi_{\bar{f}}(x)Z_{\mu}(x),
\end{split}
\end{equation}
where $\theta_{W}$ is the Weinberg angle. $g_{V}$ and $g_{A}$ are, respectively, the vector and axial-vector coupling constants.
Since the incoming $Z$-boson, which is electrically neutral, does not interact with the EM field, its wave function does not change and it is given by
\begin{equation}\label{zwave}
Z_{\mu}(x)=\frac{\varepsilon_{\mu}(q,\lambda)}{\sqrt{2QV}}e^{-iq.x},
\end{equation}
where $Q=q^{0}$ and $\varepsilon_{\mu}(q,\lambda)$ is the polarization vector of the $Z$-boson such that the summation over all three directions of polarization $\lambda$ yields $\sum_{\lambda=1}^{3}\varepsilon_{\mu}(q,\lambda)\varepsilon^{*}_{\nu}(q,\lambda)=-g_{\mu\nu}+q_{\mu}q_{\nu}/M_{Z}^{2}$, where $M_{Z}$ is the rest mass of the $Z$-boson. The outgoing neutrinos are treated as massless particles described by Dirac spinors. The outgoing charged fermions are described by the relativistic Dirac-Volkov functions normalized to the volume $V$ \cite{volkov}
\begin{equation}\label{wavefunctions}
\begin{split}
\psi_{f}(x)&=\bigg[1+\dfrac{e\slashed{k}\slashed{A}}{2(k.p_{1})}\bigg]\frac{u(p_{1},s_{1})}{\sqrt{2Q_{1}V}}\times e^{iS(q_{1},x)},\\
\psi_{\bar{f}}(x)&=\bigg[1-\dfrac{e\slashed{k}\slashed{A}}{2(k.p_{2})}\bigg]\frac{v(p_{2},s_{2})}{\sqrt{2Q_{2}V}}\times e^{-iS(q_{2},x)},
\end{split}
\end{equation}
where regarding the sign one should note that $e=-|e|$ is the charge of the electron, and
\begin{equation}
\begin{split}
S(q_{1},x)=&-q_{1}.x-\dfrac{e(a_{1}.p_{1})}{k.p_{1}}\sin(\phi)+\dfrac{e(a_{2}.p_{1})}{k.p_{1}}\cos(\phi),\\
S(q_{2},x)=&-q_{2}.x-\dfrac{e(a_{1}.p_{2})}{k.p_{2}}\sin(\phi)+\dfrac{e(a_{2}.p_{2})}{k.p_{2}}\cos(\phi).
\end{split}
\end{equation}
$u(p_{1},s_{1})$ and $v(p_{2},s_{2})$ represent the Dirac bispinors for the free charged fermion with momentum $p_{i=1,2}$ and spin $s_{i=1,2}$ satisfying, respectively, $\sum_{s_{1}}u(p_{1},s_{1})\overline{u}(p_{1},s_{1})=\slashed{p}_{1}+m_{f}$ and $\sum_{s_{2}}v(p_{2},s_{2})\overline{v}(p_{2},s_{2})=\slashed{p}_{2}-m_{f},$
where $m_{f}$ is the rest mass of the charged fermion.
The 4-vector $q_{i}(i=1,2)=p_{i}-\big[e^{2}a^{2}/2(k.p_{i})\big]k$ is the quasi-momentum  that the charged fermion acquires in the presence of the EM field. Inserting Eqs.~(\ref{zwave}) and (\ref{wavefunctions}) into Eq.~(\ref{smatrix}) and after some algebraic manipulations, we find the S-matrix to be written as
\begin{align}
S_{fi}=\dfrac{-ig}{4\cos(\theta_{W})}\frac{1}{\sqrt{8QQ_{1}Q_{2}V^{3}}}\sum_{s=-\infty}^{\infty}M^{s}_{fi}(2\pi)^{4}\delta^{4}(q_{1}+q_{2}-q-sk),
\end{align}
where $s$  is the number of exchanged photons. The quantity $M^{s}_{fi}$ is defined by
\begin{align}\label{xwithx}
M^{s}_{fi}=\bar{u}(p_{1},s_{1})\Lambda_{s}v(p_{2},s_{2})\varepsilon_{\mu}(q,\lambda),~~~~\text{with}~~\Lambda_{s}=C_{0}B_{s}(z)+ C_{1}B_{1s}(z)+C_{2}B_{2s}(z).
\end{align}
The coefficients $B_{s}(z)$, $B_{1s}(z)$ and $B_{2s}(z)$ are expressed in terms of Bessel functions by \cite{landau}
\begin{align}
\begin{split}
\begin{bmatrix}
B_{s}(z)\\
B_{1s}(z)\\
B_{2s}(z) \end{bmatrix}=\begin{bmatrix}J_{s}(z)e^{is\phi_{0}}\\
\big(J_{s+1}(z)e^{i(s+1)\phi_{0}}+J_{s-1}(z)e^{i(s-1)\phi_{0}}\big)/2\\
\big(J_{s+1}(z)e^{i(s+1)\phi_{0}}-J_{s-1}(z)e^{i(s-1)\phi_{0}}\big)/2i
 \end{bmatrix},
\end{split}
\end{align}
with $z=\sqrt{\alpha_{1}^{2}+\alpha_{2}^{2}}$ and $\phi_{0}=\arctan(\alpha_{2}/\alpha_{1})$, where 
\begin{equation}
\begin{split}
\alpha_{1}&=-e\bigg(\dfrac{a_{1}.p_{1}}{k.p_{1}}+\dfrac{a_{1}.p_{2}}{k.p_{2}}\bigg),\\
\alpha_{2}&=-e\bigg(\dfrac{a_{2}.p_{1}}{k.p{1}}+\dfrac{a_{2}.p_{2}}{k.p_{2}}\bigg).
\end{split}
\end{equation}
In Eq.~(\ref{xwithx}), the three quantities $C_{0}$, $C_{1}$, and $C_{2}$ are expressed as follows
\begin{equation}\label{constants}
\begin{split}
C_{0}&=\gamma^{\mu}(g_{V}-g_{A}\gamma_{5})+2a^{2}k^{\mu}\slashed{k}(g_{V}-g_{A}\gamma_{5})C(p_{1})C(p_{2}),\\
C_{1}&=-C(p_{2})\gamma^{\mu}(g_{V}-g_{A}\gamma_{5})\slashed{k}\slashed{a}_{1}+C(p_{1})\slashed{a}_{1}\slashed{k}\gamma^{\mu}(g_{V}-g_{A}\gamma_{5}),\\
C_{2}&=-C(p_{2})\gamma^{\mu}(g_{V}-g_{A}\gamma_{5})\slashed{k}\slashed{a}_{2}+C(p_{1})\slashed{a}_{2}\slashed{k}\gamma^{\mu}(g_{V}-g_{A}\gamma_{5}),
\end{split}
\end{equation}
where $C(p_{i})(i=1,2)=e/2 (k.p_{i})$.
To evaluate the $Z$-boson lifetime, we first evaluate the decay rate of the $Z$-boson per particle and per time into the final states, which are obtained by multiplying the squared S-matrix element by the density of final states, summing over spins of fermions and antifermions, averaging over the polarization of the incoming boson and finally dividing by the time $T$. We obtain for the decay rate of the $Z$-boson
\begin{align}\label{summed}
\Gamma(Z\rightarrow f\bar{f})=\sum_{s=-\infty}^{+\infty}\Gamma^{s},
\end{align}
where the photon-number-resolved decay rate $\Gamma^{s}$ is defined by
\begin{align}
\begin{split}
\Gamma^{s}&=\dfrac{g^{2}N_{c}}{128\cos^{2}(\theta_{W})Q}\int\dfrac{d^{3}q_{1}}{(2\pi)^{3}Q_{1}}\int\dfrac{d^{3}q_{2}}{(2\pi)^{3}Q_{2}}\\&\times(2\pi)^{4}\delta^{4}(q_{1}+q_{2}-q-sk)|\overline{M^{s}_{fi}}|^{2},
\end{split}
\end{align}
where $N_{c}$ denotes the number of colors ($N_{c}=3$ for a pair of quarks and $N_{c}=1$ otherwise) and
\begin{align}
|\overline{M^{s}_{fi}}|^{2}=\frac{1}{3}\sum_{\lambda}\sum_{s_{1},s_{2}}|M^{s}_{fi}|^{2}.
\end{align}
Performing the integration over $d^{3}q_{2}$ and using $ d^{3}q_{1}=|\textbf{q}_{1}|^{2}d|\textbf{q}_{1}|d\Omega_{f}$, the photon-number-resolved decay rate $\Gamma^{s}$ becomes
\begin{align}\label{28}
\begin{split}
\Gamma^{s}=&\dfrac{g^{2}N_{c}}{128(2\pi)^{2}\cos^{2}(\theta_{W})Q}\int\dfrac{|\textbf{q}_{1}|^{2}d|\textbf{q}_{1}|d\Omega_{f}}{Q_{1}Q_{2}}\delta(Q_{1}+Q_{2}-Q-s\omega)|\overline{M^{s}_{fi}}|^{2},
\end{split}
\end{align}
with $\textbf{q}_{1}+\textbf{q}_{2}-\textbf{q}-s\textbf{k}=0$. 
In the $Z$-boson rest frame ($Q=M_{Z}$ and $\textbf{q}=0$), the remaining integral over $d|\textbf{q}_{1}|$ can be solved by using the familiar formula \cite{greiner} 
\begin{align}\label{familiarformula}
\int dxf(x)\delta(g(x))=\dfrac{f(x)}{|g'(x)|}\bigg|_{g(x)=0}.
\end{align}
Thus, we get
\begin{align}\label{ws}
\begin{split}
\Gamma^{s}&=\dfrac{\text{G}_{\text{F}}M_{Z}N_{c}}{16\sqrt{2}(2\pi)^{2}}\int \dfrac{|\textbf{q}_{1}|^{2}d\Omega_{f}}{Q_{1}Q_{2}g'(|\textbf{q}_{1}|)}|\overline{M^{s}_{fi}}|^{2},
\end{split}
\end{align}
where we have used $g^{2}=8\text{G}_{\text{F}}M_{Z}^{2}\cos^{2}(\theta_{W})/\sqrt{2}$, with $\text{G}_{\text{F}}=(1.166~37\pm0.000~02)\times10^{-11}~\text{MeV}^{-2}$ is the Fermi coupling constant, and
\begin{align}
\begin{split}
g'(|\textbf{q}_{1}&|)=\dfrac{|\textbf{q}_{1}|}{\sqrt{|\textbf{q}_{1}|^{2}+m_{f*}^{2}}}+\dfrac{|\textbf{q}_{1}|-s\omega\cos(\theta)}{\sqrt{(s\omega)^{2}+|\textbf{q}_{1}|^{2}-2s\omega|\textbf{q}_{1}|\cos(\theta)+m_{f*}^{2}}}.
\end{split}
\end{align}
The quantity $m_{f*}=\sqrt{m_{f}^{2}-e^{2}a^{2}}$ is an effective mass of the charged fermion inside the EM field. The integral over $d\Omega_{f}$ $(d\Omega_{f}=\sin(\theta)d\theta d\varphi)$ involved in the evaluation of $\Gamma^{s}$ (\ref{ws}) is performed using the numerical integration. The coupling constants $g_{V}$ and $g_{A}$ appearing in Eq.~(\ref{constants}) are defined as
\begin{equation}
\begin{split}
Z\rightarrow l^{+}l^{-}~:~g_{V}&=-1+4\sin^{2}(\theta_{W})~;~g_{A}=1,\\
Z\rightarrow \text{up-quarks}~:~g_{V}&=1-\frac{8}{3}\sin^{2}(\theta_{W})~;~g_{A}=-1,\\
Z\rightarrow \text{down-quarks}~:~g_{V}&=-1+\frac{4}{3}\sin^{2}(\theta_{W})~;~g_{A}=1,
\end{split}
\end{equation}
and $g_{V}=g_{A}=1$ for $Z\rightarrow \text{neutrinos}$.\\
The term $|\overline{M^{s}_{fi}}|^{2}$ can be calculated as follows:
\begin{equation}
\begin{split}
|\overline{M^{s}_{fi}}|^{2}=\frac{1}{3}&\sum_{\lambda}\sum_{s_{1},s_{2}}|M^{s}_{fi}|^{2}=\frac{1}{3}\big(-g_{\mu\nu}+\frac{q_{\mu}q_{\nu}}{M_{Z}^{2}}\big)\text{Tr}\big[(\slashed{p}_{1}+m_{f})\Lambda_{s}(\slashed{p}_{2}-m_{f})\overline{\Lambda}_{s}\big],
\end{split}
\end{equation}
where
\begin{align}
\begin{split}
\overline{\Lambda}_{s}&=\gamma^{0}\Lambda_{s}^{\dagger}\gamma^{0}=\overline{C}_{0}B^{*}_{s}(z)+ \overline{C}_{1}B^{*}_{1s}(z)+\overline{C}_{2}B^{*}_{2s}(z),
\end{split}
\end{align}
and
\begin{equation}
\begin{split}
\overline{C}_{0}&=\gamma^{0}C_{0}^{\dagger}\gamma^{0}=\gamma^{\nu}(g_{V}-g_{A}\gamma_{5})+2a^{2}k^{\nu}\slashed{k}(g_{V}-g_{A}\gamma_{5})C(p_{1})C(p_{2}),\\
\overline{C}_{1}&=\gamma^{0}C_{1}^{\dagger}\gamma^{0}=-C(p_{2})\slashed{a}_{1}\slashed{k}\gamma^{\nu}(g_{V}-g_{A}\gamma_{5})+C(p_{1})\gamma^{\nu}(g_{V}-g_{A}\gamma_{5})\slashed{k}\slashed{a}_{1},\\
\overline{C}_{2}&=\gamma^{0}C_{2}^{\dagger}\gamma^{0}=-C(p_{2})\slashed{a}_{2}\slashed{k}\gamma^{\nu}(g_{V}-g_{A}\gamma_{5})+C(p_{1})\gamma^{\nu}(g_{V}-g_{A}\gamma_{5})\slashed{k}\slashed{a}_{2}.
\end{split}
\end{equation}
The trace calculations are performed with the help of FEYNCALC \cite{feyncalc}. The result obtained for $|\overline{M^{s}_{fi}}|^{2}$ is given by
\begin{align}\label{result}
\begin{split}
|\overline{M^{s}_{fi}}|^{2}=\dfrac{1}{3}\bigg[A&J_{s}^{2}(z)+BJ_{s+1}^{2}(z)+CJ_{s-1}^{2}(z)+DJ_{s}(z)J_{s+1}(z)+EJ_{s}(z)J_{s-1}(z)\\&+FJ_{s+1}(z)J_{s-1}(z)\bigg],
\end{split}
\end{align}
where the six coefficients $A$, $B$, $C$, $D$, $E$ and $F$ are explicitly expressed by
\begin{equation}
\begin{split}
A=&\frac{2}{(k.q_{1})^2 (k.q_{2})^2 M_{Z}^2}\Big[a^4 e^4 (g_{A}^2 + g_{V}^2) (k.k_{1})^2 (k.p_{1}) (k.p_{2}) - 
   2 a^2 e^2 (k.q_{1}) (k.q_{2}) (g_{A}^2 ((k_{1}.p_{2}) \\&\times(k.k_{1}) (k.p_{1}) + (k_{1}.p_{1}) (k.k_{1}) (k.p_{2}) - 
         2 (k.p_{1}) (k.p_{2}) M_{Z}^2 + (k.k_{1})^2 (m_{f}^{2} - (p_{1}.p_{2}))) \\&+ 
      g_{V}^2 ((k_{1}.p_{2}) (k.k_{1}) (k.p_{1}) + (k_{1}.p_{1}) (k.k_{1}) (k.p_{2}) - 2 (k.p_{1}) (k.p_{2}) M_{Z}^2 - 
         (k.k_{1})^2 (m_{f}^{2}\\&+ (p_{1}.p_{2}))))  
   2 (k.q_{1})^2 (k.q_{2})^2 (g_{A}^2 (2 (k_{1}.p_{1}) (k_{1}.p_{2}) + M_{Z}^2 (-3 m_{f}^{2} + (p_{1}.p_{2}))) \\&+ 
      g_{V}^2 (2 (k_{1}.p_{1}) (k_{1}.p_{2}) + M_{Z}^2 (3 m_{f}^{2} + (p_{1}.p_{2}))))\Big],      
\end{split}
\end{equation}
\begin{equation}
\begin{split}
B=&\frac{-e^2}{2(k.q_{1})^2 (k.q_{2})^2 M_{Z}^2}
  \Big[2 (-2 (a_{1}.p_{1}) (a_{1}.p_{2}) (g_{A}^2 + g_{V}^2) (k.k_{1})^2 (k.q_{1}) (k.q_{2}) + 
       a^2 (g_{A}^2 (2 (k_{1}.p_{1}) \\&\times(k.k_{1}) (k.p_{2}) (k.q_{1}) ((k.q_{1}) - (k.q_{2})) 
             2 (k_{1}.p_{2}) (k.k_{1}) (k.p_{1}) (k.q_{2}) (-(k.q_{1}) + (k.q_{2})) \\&-2 (k.k_{1})^2(k.q_{1}) (k.q_{2}) m_{f}^{2}  
             (k.p_{1}) (k.p_{2}) (k.q_{1})^2 M_{Z}^2 + 2 (k.p_{1}) (k.p_{2}) (k.q_{1}) (k.q_{2}) M_{Z}^2 \\&+ 
             (k.p_{1}) (k.p_{2}) (k.q_{2})^2 M_{Z}^2 + 2 (k.k_{1})^2 (k.q_{1}) (k.q_{2}) (p_{1}.p_{2})) + 
          g_{V}^2 (2 (k_{1}.p_{1}) (k.k_{1}) (k.p_{2})\\&\times (k.q_{1}) ((k.q_{1}) - (k.q_{2}))+ 
             2 (k_{1}.p_{2})  (k.k_{1})(k.p_{1}) (k.q_{2}) (-(k.q_{1}) + (k.q_{2})) + 2 (k.k_{1})^2 \\&\times(k.q_{1}) (k.q_{2}) m_{f}^{2} + 
             (k.p_{1}) (k.p_{2}) (k.q_{1})^2 M_{Z}^2 +2 (k.p_{1})  (k.p_{2}) (k.q_{1})(k.q_{2}) M_{Z}^2 + 
             (k.p_{1}) (k.p_{2})\\&\times (k.q_{2})^2 M_{Z}^2 + 2 (k.k_{1})^2 (k.q_{1}) (k.q_{2}) (p_{1}.p_{2})))) + 
    g_{A} g_{V} (k.k_{1}) (-(k.q_{1})^2 + (k.q_{2})^2) (p_{1}.p_{2}) \\&\times \epsilon(a_{1},a_{2},k,k_{1}) + 
    g_{A} g_{V} ((k_{1}.p_{2}) (k.k_{1}) (k.q_{2}) (-(k.q_{1}) + 3 (k.q_{2}))+ 
       2 (k.p_{2}) (-(k.q_{1}) (k.q_{2}))\\&\times (3 (k.q_{1})  + (k.q_{2})) M_{Z}^2) \epsilon(a_{1},a_{2},k,p_{1}) - 
    g_{A} g_{V} (((k_{1}.p_{1}) (k.k_{1})(k.q_{1})(3 (k.q_{1}) - (k.q_{2}))  \\&+ 
         2 (k.p_{1}) ((k.q_{1})  - (k.q_{2}))((k.q_{1}) + 3 (k.q_{2})) M_{Z}^2) \epsilon(a_{1},a_{2},k,p_{2}) + 
       (k.k_{1}) ((k.p_{2}) (-(k.q_{1})^2 \\&+ 5 (k.q_{1}) (k.q_{2}) + 2 (k.q_{2})^2) \epsilon(a_{1},a_{2},k_{1},p_{1}) + 
          (k.p_{1}) (-2 (k.q_{1})^2 - 5 (k.q_{1})(k.q_{2}) \\&+ (k.q_{2})^2)\epsilon(a_{1},a_{2},k_{1},p_{2}) + ((k.q_{1}) + (k.q_{2})) (3 (k.k_{1}) ((k.q_{1}) + (k.q_{2})) \epsilon(a_{1},a_{2},p_{1},p_{2}) \\&-
              (a_{1}.p_{2}) (k.q_{2}) \epsilon(a_{2},k,k_{1},p_{1}) + 
             (a_{1}.p_{1}) (k.q_{1}) \epsilon(a_{2},k,k_{1},p_{2}))))\Big],     
\end{split}
\end{equation}
\begin{equation}
\begin{split}
C=&\frac{-e^2}{2(k.q_{1})^2 (k.q_{2})^2 M_{Z}^2}
  \Big[2 (-2 (a_{1}.p_{1}) (a_{1}.p_{2}) (g_{A}^2 + g_{V}^2) (k.k_{1})^2 (k.q_{1}) (k.q_{2}) + 
       a^2 (g_{A}^2 (2 (k_{1}.p_{1})\\&\times (k.k_{1}) \times(k.p_{2}) (k.q_{1}) ((k.q_{1}) - (k.q_{2})) + 
             2 (k_{1}.p_{2}) (k.k_{1}) (k.p_{1}) (k.q_{2}) (-(k.q_{1}) + (k.q_{2}))\\& - 2 (k.k_{1})^2 (k.q_{1}) (k.q_{2}) m_{f}^{2} + 
             (k.p_{1}) (k.p_{2}) (k.q_{1})^2 M_{Z}^2 + 2 (k.p_{1}) (k.p_{2}) (k.q_{1}) (k.q_{2}) M_{Z}^2 \\&+ 
             (k.p_{1}) (k.p_{2})(k.q_{2})^2 M_{Z}^2 + 2 (k.k_{1})^2 (k.q_{1}) (k.q_{2}) (p_{1}.p_{2})) + 
          g_{V}^2 (2 (k_{1}.p_{1}) (k.k_{1}) (k.p_{2})\\&\times  (k.q_{1}) ((k.q_{1}) - (k.q_{2})) + 
             2 (k_{1}.p_{2}) (k.k_{1}) (k.p_{1}) (k.q_{2}) (-(k.q_{1}) + (k.q_{2})) + 2 (k.k_{1})^2 (k.q_{1})\\&\times (k.q_{2}) m_{f}^{2} + 
             (k.p_{1}) (k.p_{2}) (k.q_{1})^2 M_{Z}^2 + 2 (k.p_{1}) (k.p_{2}) (k.q_{1}) (k.q_{2}) M_{Z}^2 + 
             (k.p_{1}) (k.p_{2}) (k.q_{2})^2 \\&\times M_{Z}^2 + 2 (k.k_{1})^2 (k.q_{1}) (k.q_{2}) (p_{1}.p_{2})))) + 
    g_{A} g_{V} (k.k_{1}) ((k.q_{1}) - (k.q_{2})) ((k.q_{1}) + (k.q_{2})) \\&\times(p_{1}.p_{2}) \epsilon(a_{1},a_{2},k,k_{1}) + 
    g_{A} g_{V} ((k_{1}.p_{2}) (k.k_{1}) ((k.q_{1}) - 3 (k.q_{2})) (k.q_{2}) + 
       2 (k.p_{2}) ((k.q_{1})\\& - (k.q_{2})) (3 (k.q_{1}) + (k.q_{2})) M_{Z}^2) \epsilon(a_{1},a_{2},k,p_{1}) + 
    g_{A} g_{V} (((k_{1}.p_{1}) (k.k_{1}) (k.q_{1}) (3 (k.q_{1}) \\&- (k.q_{2})) + 
          2 (k.p_{1}) ((k.q_{1}) - (k.q_{2})) ((k.q_{1}) + 3 (k.q_{2})) M_{Z}^2) \epsilon(a_{1},a_{2},k,p_{2}) + 
       (k.k_{1}) ((k.p_{2})\\&\times (-(k.q_{1})^2 + 5 (k.q_{1}) (k.q_{2}) + 2 (k.q_{2})^2) \epsilon(a_{1},a_{2},k_{1},p_{1}) + 
          (k.p_{1}) (-2 (k.q_{1})^2 - 5 (k.q_{1}) (k.q_{2}) \\&+ (k.q_{2})^2) \epsilon(a_{1},a_{2},k_{1},p_{2}) + ((k.q_{1}) + (k.q_{2})) (3 (k.k_{1}) ((k.q_{1}) + (k.q_{2})) \epsilon(a_{1},a_{2},p_{1},p_{2}) \\&-
              (a_{1}.p_{2}) (k.q_{2}) \epsilon(a_{2},k,k_{1},p_{1}) + 
             (a_{1}.p_{1}) (k.q_{1}) \epsilon(a_{2},k,k_{1},p_{2}))))\Big],
\end{split}
\end{equation}
\begin{equation}
\begin{split}
D=&\frac{-2e}{(k.q_{1})^2 (k.q_{2})^2 M_{Z}^2}
  \Big[(g_{A}^2 + 
       g_{V}^2) ((k.k_{1}) (-2 (k.q_{1}) (k.q_{2}) ((a_{1}.p_{2}) (k_{1}.p_{1}) (k.q_{1}) + (a_{1}.p_{1}) \\&\times(k_{1}.p_{2}) (k.q_{2})) + 
          a^2 e^2 (k.k_{1}) ((a_{1}.p_{1}) (k.p_{2}) (k.q_{1}) + (a_{1}.p_{2}) (k.p_{1}) (k.q_{2}))) + (-(a_{1}.p_{2}) \\&\times(k.p_{1}) + 
          (a_{1}.p_{1}) (k.p_{2})) (k.q_{1}) ((k.q_{1}) - (k.q_{2})) (k.q_{2}) M_{Z}^2) + 
    g_{A} g_{V} ((k.q_{1}) - (k.q_{2})) (a^2 e^2 \\&\times(k.k_{1}) (k.p_{2}) - 2 (k_{1}.p_{2}) (k.q_{1})(k.q_{2})) \epsilon(a_{2},k,k_{1},p_{1}) + 
    g_{A} g_{V} ((k.q_{1}) - (k.q_{2})) (a^2 e^2 (k.k_{1})\\&\times (k.p_{1})- 2 (k_{1}.p_{1}) (k.q_{1}) (k.q_{2})) \epsilon(a_{2},k,k_{1},p_{2}) + 
    g_{A} g_{V} ((k.q_{1}) + 
       (k.q_{2})) (-(a^2 e^2 (k.k_{1})^2  \\&+ 4 (k.q_{1}) (k.q_{2}) M_{Z}^2) \epsilon(a_{2},k,p_{1},p_{2})+ 
       2 (k.k_{1}) (k.q_{1}) (k.q_{2}) \epsilon(a_{2},k_{1},p_{1},p_{2}))\Big],
\end{split}
\end{equation}
\begin{equation}
\begin{split}
E=&\frac{-2e}{(k.q_{1})^2 (k.q_{2})^2 M_{Z}^2}
  \Big[(g_{A}^2 + 
       g_{V}^2) ((k.k_{1}) (-2 (k.q_{1}) (k.q_{2}) ((a_{1}.p_{2}) (k_{1}.p_{1}) (k.q_{1}) + (a_{1}.p_{1}) \\&\times(k_{1}.p_{2}) (k.q_{2})) + 
          a^2 e^2 (k.k_{1}) ((a_{1}.p_{1}) (k.p_{2}) (k.q_{1}) + (a_{1}.p_{2}) (k.p_{1}) (k.q_{2}))) + (-(a_{1}.p_{2}) (k.p_{1}) \\&+ 
          (a_{1}.p_{1}) (k.p_{2})) (k.q_{1}) ((k.q_{1}) - (k.q_{2})) (k.q_{2}) M_{Z}^2) + 
    g_{A} g_{V} (((k.q_{1}) - (k.q_{2})) (-a^2 e^2 (k.k_{1})\\& \times (k.p_{2})+ 2 (k_{1}.p_{2}) (k.q_{1}) (k.q_{2})) \epsilon(a_{2},k,k_{1},p_{1}) + ((k.q_{1}) - (k.q_{2})) (-a^2 e^2 (k.k_{1}) (k.p_{1})\\& + 
          2 (k_{1}.p_{1}) (k.q_{1}) (k.q_{2})) \epsilon(a_{2},k,k_{1},p_{2}) + ((k.q_{1}) + 
          (k.q_{2})) ((a^2 e^2 (k.k_{1})^2 + 4 (k.q_{1}) (k.q_{2}) M_{Z}^2) \\&\times\epsilon(a_{2},k,p_{1},p_{2}) - 
          2 (k.k_{1}) (k.q_{1}) (k.q_{2}) \epsilon(a_{2},k_{1},p_{1},p_{2})))\Big],
\end{split}
\end{equation}
\begin{equation}
\begin{split}
F&=\frac{4 e^2  (g_{A}^2 + g_{V}^2)(k.k_{1})^2(a_{1}.p_{1}) (a_{1}.p_{2})  }{(k.q_{1}) (k.q_{2}) M_{Z}^2},
\end{split}
\end{equation}
where, for all 4-vectors $a, b, c$ and $d$, we have
\begin{equation}
\epsilon(a,b,c,d)=\epsilon_{\mu\nu\rho\sigma}a^{\mu}b^{\nu}c^{\rho}d^{\sigma}.
\end{equation}
As we can see, in the two coefficients $A$ and $F$, there is no appearance of the antisymmetric tensors $\epsilon_{\mu\nu\rho\sigma}$. This obviously implies that they were completely contracted. But the other coefficients $B$, $C$, $D$
and $E$ contained various noncontracted tensors. In order to evaluate these tensors, we remind here that we have used the Grozin convention
\begin{equation}
\epsilon_{0123}=1.
\end{equation}
To be more clear, we give here how we calculated one of these tensors, for example in $B$ and $C$:
\begin{equation}
\begin{split}
\epsilon(a_{1},a_{2},p_{1},p_{2})&=\epsilon_{\mu\nu\rho\sigma}a_{1}^{\mu}a_{2}^{\nu}p_{1}^{\rho}p_{2}^{\sigma},\\
&=|\text{\textbf{a}}|^{2}\big[\epsilon_{1203}p_{1}^{0}p_{2}^{3}+\epsilon_{1230}p_{1}^{3}p_{2}^{0}\big],\\
&=|\text{\textbf{a}}|^{2}\Big[\Big(Q_{1}+\frac{e^{2}a^{2}\omega}{2(k.q_{1})}\Big)\Big(s\omega-|\textbf{q}_{1}|\cos(\theta)+\frac{e^{2}a^{2}\omega}{2(k.q_{2})}\Big)\\&~~~-\Big(|\textbf{q}_{1}|\cos(\theta)+\frac{e^{2}a^{2}\omega}{2(k.q_{1})}\Big)\Big(Q_{2}+\frac{e^{2}a^{2}\omega}{2(k.q_{2})}\Big)\Big].
\end{split}
\end{equation}
We return to the lifetime of the $Z$-boson, which is defined by
\begin{align}
\tau_{Z}=\dfrac{1}{\Gamma_{\text{tot}}},
\end{align}
where $\Gamma_{\text{tot}}$ is the total decay rate of the $Z$-boson in the laser field given by
\begin{equation}
\Gamma_{\text{tot}}=\Gamma(Z\rightarrow \text{hadrons})+\Gamma(Z\rightarrow l^{+}l^{-})+\Gamma_{\text{inv}},
\end{equation}
where $\Gamma(Z\rightarrow \text{hadrons})=\Gamma(Z\rightarrow \text{up-quarks})+\Gamma(Z\rightarrow \text{down-quarks}),$
and $\Gamma_{\text{inv}}=\Gamma(Z\rightarrow \text{neutrinos}).$\\
The branching ratio (BR) is the ratio between each partial decay rate and the total decay rate of the $Z$. It refers, in particle physics, to the likelihood that a particle will decay to a particular mode out of all possible decay modes. The sum of the branching ratios of all decay modes of a particle is therefore by definition equal to $1$ (or $100\%$).
We define the three (BRs) of the different $Z$-boson decay modes as follows
\begin{eqnarray}
\text{BR}(Z\rightarrow \text{hadrons})&=&\frac{\Gamma(Z\rightarrow \text{hadrons})}{\Gamma_{\text{tot}}},\label{brhadrons}\\
\text{BR}(Z\rightarrow l^{+}l^{-})&=&\frac{\Gamma(Z\rightarrow l^{+}l^{-})}{\Gamma_{\text{tot}}},\label{brleptons}\\
\text{BR}_{\text{inv}}(Z\rightarrow \text{neutrinos})&=&\frac{\Gamma_{\text{inv}}(Z\rightarrow \text{neutrinos})}{\Gamma_{\text{tot}}}.\label{brinv}
\end{eqnarray}
Their experimental values in the absence of the laser field are \cite{pdg2020} 
\begin{equation}
\begin{split}
\text{BR}(Z\rightarrow \text{hadrons})&=(69.911\pm0.056)\%,\\
\text{BR}(Z\rightarrow l^{+}l^{-})&=(10.099\pm 0.011)\%,\\
\text{BR}_{\text{inv}}(Z\rightarrow \text{neutrinos})&=(20.000\pm0.055)\%.
\end{split}
\end{equation}
\section{Results and Discussion}
In this section, we will try to discuss and analyze the numerical results obtained regarding the decay of the $Z$-boson in the presence of a circularly polarized EM field. The important point to be addressed is how an EM field can influence the BRs or contribute to their enhancement or suppression. In the $Z$-boson decay, we have, as we saw in the previous section, three BRs. One for the hadronic channel, the second for the charged leptonic channel, and the other for the invisible channel. We will present in the following the effect of the laser intensity on each of these ratios. The hadronic $\text{BR}(Z\rightarrow \text{hadrons})$ is the largest branching ratio among these ratios, and this means that the decay of the $Z$-boson into a pair of quarks in the absence of an EM field has a great probability compared to other pairs.
\begin{figure}[hbtp]
 \centering
 \includegraphics[scale=0.7]{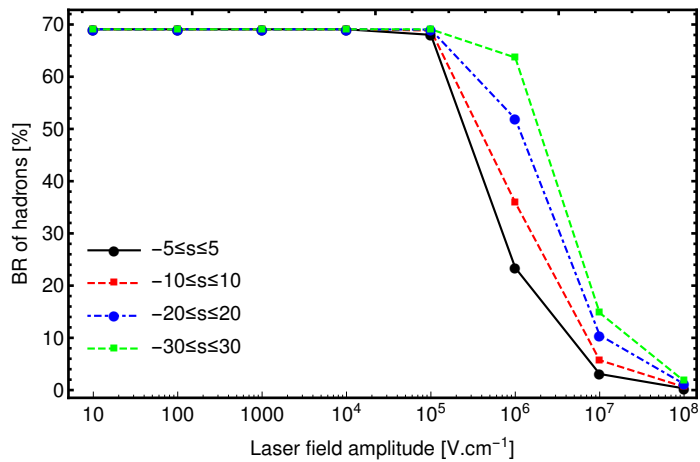}
 \caption{The behavior of the BR (\ref{brhadrons}) of the hadronic decay mode as a function of the laser field amplitude for different numbers of photons exchanged. The frequency of
laser field is $\hbar\omega=1.17~\text{eV}$.}\label{fig1}
 \end{figure}
 \begin{figure}[hbtp]
 \centering
 \includegraphics[scale=0.7]{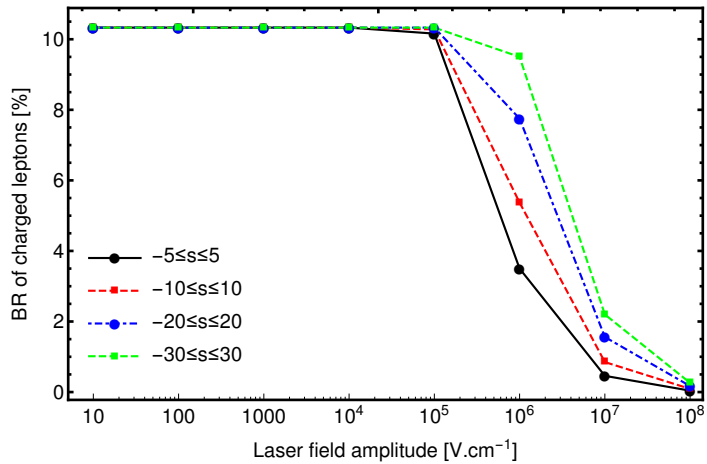}
 \caption{The behavior of the BR (\ref{brleptons}) of the charged leptonic decay mode as a function of the laser field amplitude for different numbers of photons exchanged. The frequency of
laser field is $\hbar\omega=1.17~\text{eV}$.}\label{fig2}
 \end{figure}
In Figs.~\ref{fig1} and \ref{fig2}, we represent the variations of $\text{BR}(Z\rightarrow \text{hadrons})$ (\ref{brhadrons}) and $\text{BR}(Z\rightarrow l^{+}l^{-})$ (\ref{brleptons}) in terms of the laser field amplitude for different numbers of exchanged photons. We note that all the equivalent curves for every given number of exchanged photons $s$ start from the value of the BRs in the absence of the EM field ($\sim69\%$ in Fig~\ref{fig1} and $\sim10\%$ in Fig~\ref{fig2}), and remain constant and identical in the interval of intensities between $10$ and
 $10^{5}~\text{V/cm}$. At this last value, curves begin to differ and disperse. In other words, the two BRs were not affected at all by the EM field at weak intensities, regardless of the number of exchanged photons. But it is clearly evident that, from the intensity $10^{5}~\text{V/cm}$, the two BRs begin to decrease according to the number of exchanged photons $s$, where as $s$ increases, the degree of decline decreases, until both BRs become almost non-existent at the intensity $10^{8}~\text{V/cm}$. Based on the foregoing, it can be said that the EM field at high intensities has greatly contributed to the suppression of both hadronic and charged leptonic channels. They became, after being favored in the absence of an EM field, almost forbidden due to the presence of the EM field. 
 \begin{figure}[hbtp]
 \centering
 \includegraphics[scale=0.7]{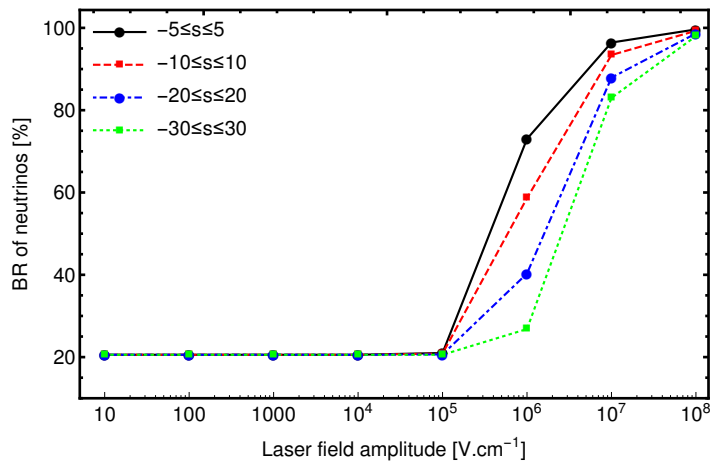}
 \caption{The behavior of the invisible BR (\ref{brinv}) as a function of the laser field amplitude for different numbers of photons exchanged. The frequency of laser field is $\hbar\omega=1.17~\text{eV}$.}\label{fig3}
 \end{figure}
On the other hand, regarding the invisible $\text{BR}_{\text{inv}}(Z\rightarrow \text{neutrinos})$, the matter is completely different. Figure \ref{fig3} depicts the behavior of the invisible BR (\ref{brinv}) versus the laser field amplitude for different numbers of photons exchanged. We notice that the invisible BR begins at its value in the absence of an EM field $(20\%)$ and remains constant up to intensity $10^{5}~\text{V/cm}$, where it rises with increasing intensity according to the number of exchanged photons $s$, until it reaches about $99\%$ at the intensity $10^{8}~\text{V/cm}$. This is normal and expected based on the behavior of the previous two BRs. Thus, the decrease  in the hadronic and charged leptonic BRs must be accompanied or compensated by the increase in the invisible BR. The suppression of hadronic and charged leptonic decay modes is offset on the other hand by the enhancement of the invisible decay mode.  
The reader may reasonably wonder, how the EM field affects the $Z$-boson decay into neutrinos even though neither the $Z$-boson nor the neutrinos interact with the EM field, so we answer by saying that the EM field does not directly affect this reaction, but rather affects in reality the decay of the $Z$-boson into charged fermions where, as we saw above, it suppresses the hadronic and charged leptonic decay modes, and thus the only option left for the $Z$-boson is to invisibly decay into neutrinos. This result related to the effect of the EM field on the hadronic and charged leptonic BRs can be reasonably explicated by reference to the effective mass quantity $m_{f*}$ acquired by the final charged fermions within the EM field. As this quantity is expressed $\big(~m_{f*}=\sqrt{m_{f}^{2}-e^{2}a^{2}}=\sqrt{m_{f}^{2}+e^{2}\mathcal{E}_{0}^{2}/\omega^{2}}~\big)$, it appears that the mass acquired by one of the final charged fermions increases with the increase in the laser field amplitude $\mathcal{E}_{0}$ until it becomes, at a specific strong intensity, larger than the mass of the boson $Z$ (for example, the effective mass values for each type of charged fermions, at the intensity $10^{16}~\text{V/cm}$ and frequency $\hbar\omega=1.17~\text{eV}$, are equal and about $168$ GeV). It is then, according to the energy-conservation law, impossible for the $Z$-boson to decay into a pair of charged fermions in the presence of a superstrong EM field. We give here the values of the different BRs at the intensity $10^{16}~\text{V/cm}$ and frequency $\hbar\omega=1.17~\text{eV}$:  $\text{BR}(Z\rightarrow \text{hadrons})=1.2781\times 10^{-7}\%$, $\text{BR}(Z\rightarrow l^{+}l^{-})=1.90968\times 10^{-8}\%$ and $\text{BR}_{\text{inv}}(Z\rightarrow \text{neutrinos})=100\%$.
The effect of laser on branching ratios has been sufficiently discussed, we would like here to study, very briefly, the effect of laser intensity and frequency on the $Z$-boson lifetime. Figure \ref{fig4} illustrates the typical behavior of the $Z$-boson lifetime in the rest frame of $Z$ for the laser frequency $\hbar\omega=1.17~\text{eV}$ and for different numbers of photons exchanged. As we can see from this figure, it seems clear that as the number of exchanged photons increases, the effect of the laser on the lifetime diminishes until it becomes zero when we reach $-650\leq s\leq+650$ exchanged photons which presents the cutoff number in the case of intensity $10^{7}~\text{V/cm}$ and frequency $\hbar\omega=1.17~\text{eV}$. This result confirms exactly what we have obtained recently in the case of the pion lifetime \cite{mouslih}. To examine the effect of laser frequency on $Z$-boson lifetime, we plot, in Fig.~\ref{fig5}, the changes in lifetime as a function of laser field amplitude for two different frequencies (Nd:YAG laser: $\hbar\omega=1.17~\text{eV}$ and $\text{CO}_{2}$ laser: $\hbar\omega=0.117~\text{eV}$).
\begin{figure}[hbtp]
\centering
\includegraphics[scale=0.7]{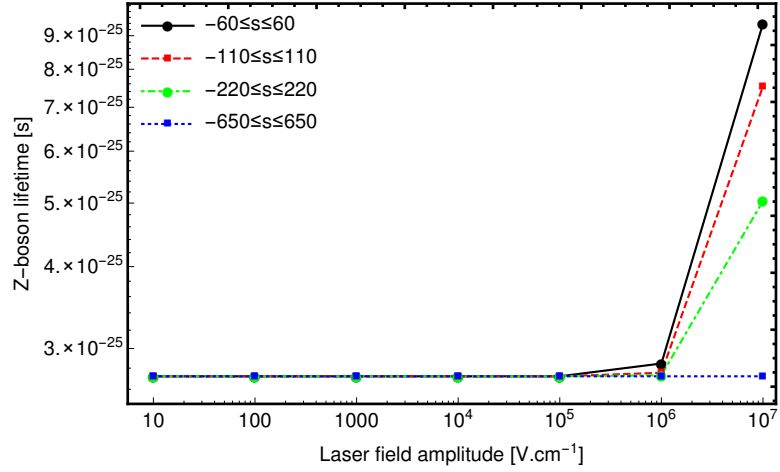}
\caption{The laser-modified $Z$-boson lifetime as a function of the laser field amplitude for various numbers of photons exchanged. The frequency of the laser field is $\hbar\omega=1.17~\text{eV}$.}\label{fig4}
\end{figure}
\begin{figure}[hbtp]
\centering
\includegraphics[scale=0.7]{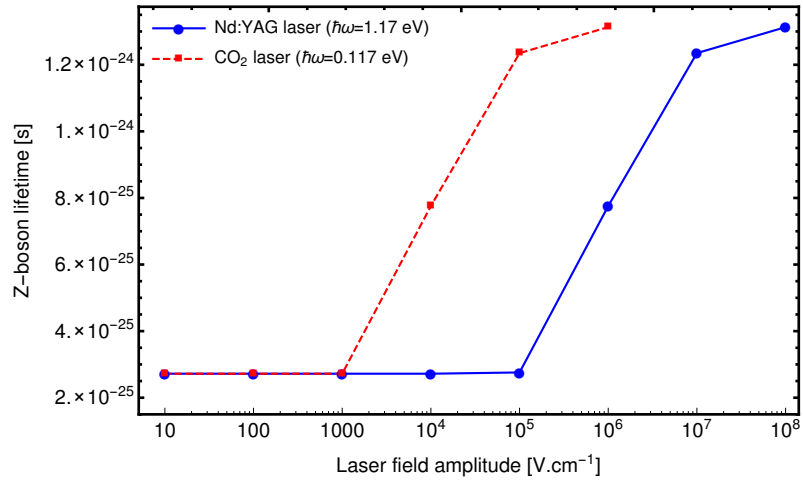}
\caption{The variations of laser-modified $Z$-boson lifetime as a function of the laser field amplitude for a Nd:YAG laser ($\hbar\omega=1.17~\text{eV}$) and a $\text{CO}_{2}$ laser ($\hbar\omega=0.117~\text{eV}$). The number of exchanged photons is taken as $-10\leq s\leq+10$.}\label{fig5}
\end{figure}
Apparently, at high frequencies, the effect of the laser on the lifetime diminishes with a similar behavior to the case of the pion and muon lifetimes \cite{mouslih,liu}.
\section{Conclusion}
Analytical calculations have been performed for the $Z$-boson decay in the presence of a circularly polarized laser field. It has been shown, theoretically, that the branching ratio of the invisible $Z$-boson decay mode can be enhanced by applying suitable laser fields. It is therefore time for experimentalists to take advantage of the powerful laser and consider it as a proposed technology allowing them to control branching ratios. Besides this advantage, it has been shown that the effect of powerful lasers on the lifetime gives the same results as those of the pion and muon decays.
We hope that this work will pave the way for other future works.

\end{document}